\documentclass[hidelinks,report,12pt,onecolumn]{IEEEtran}
\usepackage{cite}
\usepackage{amsmath}
\usepackage{array}
\usepackage{color}

\ifCLASSOPTIONcompsoc
  \usepackage[caption=false,font=normalsize,labelfont=sf,textfont=sf]{subfig}
\else
  \usepackage[caption=false,font=footnotesize]{subfig}

\usepackage{dblfloatfix}
\usepackage{url}

\usepackage{color}
\usepackage{amsthm}
\usepackage{amsmath}
\usepackage{amssymb}
\usepackage{amsfonts}
\usepackage{csquotes}
\usepackage{mathrsfs}
\usepackage{mathtools}

\usepackage{graphicx}
\usepackage{ragged2e}
\usepackage{footnote}
\usepackage[ruled,linesnumbered]{algorithm2e}
\usepackage{algpseudocode}
\usepackage[normalem]{ulem}
\usepackage{dirtytalk}
\usepackage{enumitem}	% For modifity left margins % http://tex.stackexchange.com/questions/170525/itemize-left-margin
\usepackage{verbatim}
\usepackage{booktabs}
\usepackage{float}

\usepackage[T1]{fontenc}
\usepackage{authblk}

\usepackage[pdfpagelabels,hypertexnames=false,breaklinks=true,bookmarksopen=true,bookmarksopenlevel=2]{hyperref}

\renewcommand{\IEEEQED}

\usepackage{amsmath}
\DeclareMathOperator*{\argmax}{arg\,max}

\usepackage{hyperref}

\makeatletter
\let\original@algocf@latexcaption\algocf@latexcaption
\long\def\algocf@latexcaption#1[#2]{%
	\@ifundefined{NR@gettitle}{%
		\def\@currentlabelname{#2}%
	}{%
		\NR@gettitle{#2}%
	}%
	\original@algocf@latexcaption{#1}[{#2}]%
}
\makeatother

\begin{document}
%
% paper title
% Titles are generally capitalized except for words such as a, an, and, as,
% at, but, by, for, in, nor, of, on, or, the, to and up, which are usually
% not capitalized unless they are the first or last word of the title.
% Linebreaks \\ can be used within to get better formatting as desired.
% Do not put math or special symbols in the title.
\title{Accurate Link Lifetime Computation in\\ Autonomous Airborne UAV Networks}

\author{Shivam~Garg\thanks{S. Garg is with the Computational Science Research Center (CSRC) at San Diego State University (SDSU), CA, USA, sgarg@sdsu.edu.}, \and Alexander~Ihler\thanks{A. Ihler is with the Donald Bren Hall School of Information and Computer Sciences, University of California, Irvine (UCI), CA, USA, ihler@ics.uci.edu.}, \and and Sunil~Kumar\thanks{S. Kumar is with the Electrical and Computer Engineering Department, SDSU, San Diego, CA, USA, skumar@sdsu.edu.}}

% make the title area
\maketitle
\setlength{\parindent}{2em}
\setlength{\parskip}{0.5em}
\linespread{1.3}

\begin{abstract}

An autonomous airborne network (AN) consists of multiple unmanned aerial vehicles (UAVs), which can self-configure to provide seamless, low-cost and secure connectivity. AN is preferred for applications in civilian and military sectors because it can improve the network reliability and fault tolerance, reduce mission completion time through collaboration, and adapt to dynamic mission requirements. However, facilitating seamless communication in such ANs is a challenging task due to their fast node mobility, which results in frequent link disruptions. Many existing AN-specific mobility-aware schemes restrictively assume that UAVs fly in straight lines, to reduce the high uncertainty in the mobility pattern and simplify the calculation of link lifetime (\textit{LLT}). Here, \textit{LLT} represents the duration after which the link between a node pair terminates. However, the application of such schemes is severely limited, which makes them unsuitable for practical autonomous ANs.

In this report, a mathematical framework is described to accurately compute the \textit{LLT} value for a UAV node pair, where each node flies independently in a randomly selected smooth trajectory. In addition, the impact of random trajectory changes on \textit{LLT} accuracy is also discussed.
\end{abstract}


% For peerreview papers, this IEEEtran command inserts a page break and
% creates the second title. It will be ignored for other modes.
%\IEEEpeerreviewmaketitle

% needed in second column of first page if using \IEEEpubid
%\IEEEpubidadjcol

\section{Introduction}
An autonomous airborne network (AN) can self-configure to provide a seamless, low-cost and secure connectivity to a multitude of devices with dynamic quality of service (QoS) requirements \cite{[R1],[R44],[R7],AI-EnabledRouting,Ref1,Ref0}. Such ANs consist of multiple, cooperative manned and unmanned aerial vehicles (UAVs), which use flight-to-flight communication to reduce latency in information sharing, and provide scalability \cite{[R1],[R2],[R44],Ref1,Ref0}. Some existing applications of AN in civilian and military sectors are environmental sensing and disaster management, traffic and urban monitoring, search and rescue operations, patrolling, and relaying networks \cite{[R1],[R2],[R3],[R4],Ref0,Ref1,AI-EnabledRouting}. However, designing a robust communication mechanism for an autonomous AN is a challenging task due to the high node mobility, which results in frequent link disruptions \cite{[R1],AI-EnabledRouting,Ref1,Ref0,[R5],[R27], [R25],[R22],[R4],[R3],[R2]}. Moreover, the existing ground-based routing protocols are not suitable for AN because the mobility pattern and speed of an airborne node are significantly different than ground-based mobile node \cite{AI-EnabledRouting,Ref1,Ref0,[R5], [R6],[R11],[R27],[R25],[R22],[R4],[R3],[R2]}. Note that the design and performance of a routing scheme depends on the underlying mobility model \cite{AI-EnabledRouting,Ref1,Ref0,[R1],[R44],[R7]}. 

A number of AN-specific mobility-aware routing protocols \cite{[R22],[R25],[R33],Gauss3D} have been proposed in the literature lately. However, they restrictively assume that UAVs fly in straight lines, to reduce the uncertainty in the mobility pattern and simplify the calculation of link lifetime (\textit{LLT}). Here, \textit{LLT} represents the duration for which a link between two nodes remains connected. Therefore, the application of these routing schemes is severely limited. Some mobility-model-agnostic routing schemes (e.g., \cite{LocalRelay,StochasticMultipathRouting,QMR}) assume that a node can obtain the recent position of the destination node by querying a central or regional location server, and use a store-carry-forward approach to deliver the packets to the destination node. However, they are not suitable for a distributed network with no central or regional location servers \cite{AI-EnabledRouting,Ref0,Ref1}. Some other routing schemes (e.g., \cite{DRLforAN, PARRoT, [R26],[R27],[R9]}) focus on predicting the future node trajectories and/or link disruptions. However, they are not reliable for autonomous ANs, where each node flies independently on a randomly selected smooth trajectory, which results in high uncertainty in mobility patterns that cannot be known a priori \cite{Ref1, [R29]}. 

To address these discussed issues, we propose a mathematical framework to compute the \textit{LLT} value of a UAV node pair, where each node flies in a randomly selected smooth trajectory. Since the future node trajectories are not known, a node cannot accurately predict the complete \textit{LLT} value at the time of link establishment with its 1-hop neighbor node \cite{Ref1,[R29]}. Therefore, a node recomputes its \textit{LLT} value using our proposed mechanism, whenever either node of the node pair changes its trajectory. Then, a node can select the longest-lasting route $R^{*}$ using the \textit{LLT} values of the links in the network as,
\begin{equation}\label{RLT}
    R^{*} = \argmax_{Route~R} {\bigg(\min_{Link~l\in R}{\big(~LLT_l\big)}\bigg)}.
\end{equation}

%Note that the words “node”, “airborne node”, and “UAV” are used interchangeably in this report.

\noindent\textbf{Report Organization:} This report is organized in four sections. An overview of the existing approaches that use a stable and longer-lasting route is discussed in Section \ref{RelatedWork}. Our proposed mathematical framework to compute \textit{LLT} is discussed in Section \ref{LLTComputation}, followed by conclusion in Section \ref{Conclusion}.

\section{Related Work}
\label{RelatedWork}
We first discuss the use of \textit{LLT} in the routing protocols, followed by the mobility models used for airborne node.

Most of the existing routing schemes (e.g., \cite{[R22],[R23],[R24],[R25],[R26],[R31]}) compute the link stability using the link and node statistics, such as signal-to-noise (SNR) ratio, received signal strength indicator (RSSI), variance in the node distance, past \textit{LLT} values and the number of acknowledgement packets received during an interval. On the other hand, some other schemes (e.g., \cite{[R32],[R33],Gauss3D}) propose a mathematical formulation to compute the \textit{LLT} value by using the characteristics of underlying mobility model. Both approaches are discussed below.

The AN topology evolves with time, which makes the identification of unbroken routes a challenging task \cite{[R22],[R23],[R24]}. A transmitter node retransmits a unicast packet seven times in the CSMA/CA (carrier-sense multiple access with collision avoidance) MAC (medium access control) protocol before recognizing a link break, which reduces the channel utilization and increases the queuing delay for the remaining packets at the node \cite{[R23]}. In addition, frequent topology changes can increase congestion in the network \cite{[R23]}, which results in the packet drop due to buffer overflow and degrades the flow QoS \cite{[R22]}. 

To prevent the packet transmission over broken routes, each node in \cite{[R23],[R24]} includes the GPS locations of itself and its 1-hop neighbor nodes in its control messages in order to create a cartography of the network at each node. Based on these locations, source node selects a route such that links do not break before the reception of new control messages. However, the control packet lengths in \cite{[R23],[R24]} increases significantly in a dense network, which results in a higher control overhead and packet collision probability. In addition, periodic reconstruction of the cartography increases the computational overhead at each node, which reduces its residual battery life.

The shortest-hop routing schemes generally select the edge nodes. Consequently, the signal strength at the receiver node is minimized, which increases the packet loss ratio. To address this issue, the routing scheme in \cite{[R25]} differentiates links based on their RSSI values using Chebyshev inequality and selects stable links with a lower variance in RSSI values. However, an accurate computation of RSSI values is difficult due to high interference from neighbor nodes in a dense network \cite{[R26]}.

A mobility-aware route selection scheme is proposed in \cite{[R22]}, where the link stability is determined by the variance in the distance values (computed using GPS) of a node pair. Node pairs with a smaller variance in their distance values are expected to remain in the range of each other for a longer duration. However, \cite{[R22]} fails to select links where the UAVs come closer because of the high variance in their monotonically decreasing distance values.

The scheme discussed in \cite{[R31]} uses the distribution of the past \textit{LLT} and the current link age (i.e., the time duration since the node pair was connected until the current timestamp) to estimate the residual \textit{LLT} (i.e., the duration from the current timestamp after which the link between the node pair would expire). However, \cite{[R31]} does not consider the effect of a trajectory change, which can result in an inaccurate \textit{LLT} computation \cite{[R29]}.

Mathematical formulation is proposed in \cite{[R32],[R33],Gauss3D} to compute \textit{LLT} for a node pair using their speed, directions of movement and current coordinates. However, these formulations are limited to ground vehicles and cannot be used for the airborne node, where the direction of movement can continuously change.

\subsection{Discussion of Existing Mobility Models}
\label{MobilityModels}
The performance of a routing protocol depends on the node distribution and their connectivity \cite{[R4],[R12]}. The node connectivity, in turn, depends on the node trajectory, and thereby, on the underlying mobility model \cite{[R4],[R12]}. Therefore, designing and evaluating routing protocols for ANs require the use of mobility models, which can produce realistic node movements \cite{[R1],[R4],[R12],[R27]}. The existing mobility models for ANs are summarized below.

Several mobility models, including random walk, random waypoint, and Manhattan grid, have been proposed for mobile networks in the literature \cite{[R1],[R2],[R11]}. However, they assume node movement in 2D space, and therefore, are not suitable for ANs \cite{[R1],[R2],[R3],[R4],[R6],[R7],[R8],[R9],[R10],[R11]}. Further, a fast-moving airborne node follows a smooth trajectory due to its aerodynamic constraints \cite{[R1],[R2],[R7]}, and cannot abruptly change its direction as its movement depends on its previous locations, speed and heading direction \cite{[R5],[R7],[R12]}. Therefore, time-based mobility models are preferred for ANs where the node movement is controlled under mathematical equations \cite{[R1],[R5],[R11],[R27]}. The paparazzi model \cite{[R6]}, smooth-turn (ST) model \cite{[R7]}, semi-random circular movement (SRCM) \cite{[R8]}, and Gauss-Markov model \cite{[R9],[R10]} are time-based mobility models discussed in the literature. 

In this report, we use ST mobility model in which each node independently selects a center and radius based on its past trajectory and rotates around the center in the clockwise (or counter-clockwise) direction for a randomly selected duration, called Wait Time, which results in a smooth realistic trajectory \cite{[R1],[R7]}. Other time-based mobility models are ill-suited because they generate limited mobility patterns. For example, node movement in the paparazzi model is limited to eight possible trajectories \cite{[R1],[R2]}; nodes rotate around a static common center in SRCM model \cite{[R1],[R2]}; and Gauss-Markov model cannot reproduce typical UAV turns \cite{[R1],[R2]}.

In addition, the above-mentioned mobility models use either a reflection boundary model in which UAV makes a sharp $180^0$ turn at the boundary or wrap-around boundary model which causes sudden node appearances and disappearances \cite{[R11]}. To address this issue, we use a buffer boundary model \cite{[R11]}, which forces a UAV to select its new direction such that it does not go out of the boundary, while complying with its aerodynamics. Possible directions for a UAV to fly are \textit{clockwise}, \textit{counter-clockwise} and \textit{straight}. Note that a UAV does not change its direction from clockwise to counter-clockwise and vice-versa without flying in the straight direction for at least a small duration, in order to maintain its stability \cite{[R12]}. 

%Moreover, the turn radius \textit{R} of an airborne node depends on its velocity \textit{V} as, $R=\frac{V}{\omega}$, where $\omega$ is the angular velocity. The Wait Time duration for which a node maintains its current trajectory is uniformly selected from the range [minWaitTime, maxWaitTime], which corresponds to the minimum and maximum duration for which the node stays on its current trajectory.

\section{Computation of Link Lifetime (LLT)}
\label{LLTComputation}
We consider a UAV network, where each UAV includes its trajectory information in its Hello message, which is broadcast periodically to its 1-hop neighbors. Here, trajectory information includes the GPS location, movement state (i.e., clockwise, counter-clockwise or straight), current center, radius, and slope (if UAV is moving in a straight direction). Alternatively, a UAV can compute its trajectory, center, radius and movement state by using its three consecutive GPS locations. Its speed can be computed using the distance travelled (or angular displacement) during a Hello interval, when the UAV goes straight (or travels on a curved trajectory). We also assume that each UAV flies at a unique altitude, to prevent node collisions. 

When the current location and trajectory details of a UAV pair are known, the corresponding \textit{LLT} is computed using the following steps:\\
\textbf{Step 1:} Find the coordinates of the future location for both UAVs at time \textit{t}.

\noindent\textbf{Step 2:} The link between a UAV pair breaks down when their distance exceeds the node transmission range. Hence, an equation with one unknown variable \textit{t} is computed using, 
\begin{equation}\label{Eq2}
    Distance~between~UAVs~at~time~t~>~Transmission~Range.
\end{equation}

\noindent\textbf{Step 3:} Select the root which best approximates \eqref{Eq2}.

Note that two UAVs establish a link at the \textit{Link Establishment Time}, when they exchange their Hello packets for the first time. The link between the UAV pair terminates, when they move out of each other’s transmission range. This time is called \textit{Link Termination Time}. Therefore, \textit{LLT} is computed as,
\begin{equation}\label{Eq3}
    LLT~=~Link~Termination~Time~-~Link~Establishment~Time
\end{equation}

Based on the possible movement states for a UAV (which are clockwise, counter-clockwise and straight), the following three cases are possible for a UAV pair. The \textit{LLT} computation for each case is different as discussed below.\\

\begin{figure}[!th]
	\centering
	\includegraphics[width=0.99\linewidth]{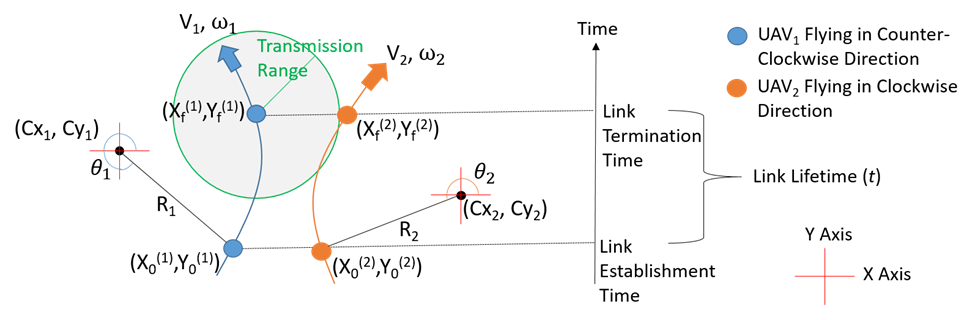}
	\caption {\textit{LLT} computation when both UAVs fly in a curve.}
	\label{Fig1}
\end{figure}

\noindent\textbf{Case A. Both UAVs fly in a curve}\par
For a $UAV_1$ and $UAV_2$ pair in Fig. \ref{Fig1}, assume their current GPS locations are ($X_{0}^{(1)}$,$Y_{0}^{(1)}$) and ($X_{0}^{(2)}$,$Y_{0}^{(2)}$), the altitudes are $Z_{0}^{(1)}$ and $Z_{0}^{(2)}$, the current trajectory centers are ($Cx_1$,$Cy_1$) and ($Cx_2$,$Cy_2$), current radii are $R_1$ and $R_2$, velocities are $V_1$ and $V_2$, and the movement directions are $Dir_1$ and $Dir_2$, respectively. Here, \textit{Dir} is -1 for clockwise direction and +1 for counter-clockwise direction.

The angular velocity for the UAV pair is computed as, $\omega_1=\frac{V_1*Dir_1}{R_1}$ and $\omega_2=\frac{V_2*Dir_2}{R_2}$. The initial displacement at the \textit{Link Establishment Time} for each UAV is computed as, $\theta_1=tan^{-1}\bigg(\frac{Y_0^{(1)}-Cy_1}{X_0^{(1)}-Cx_1}\bigg)$ and $\theta_2=tan^{-1}\bigg(\frac{Y_0^{(2)}-Cy_2}{X_0^{(2)}-Cx_2}\bigg)$. Hence, the future location coordinates at time \textit{t} for both UAVs are,

$\bigg(X_f^{(1)},Y_f^{(1)},Z_f^{(1)}\bigg) = \big(Cx_1 + R_1 cos(\theta_1+\omega_1 t),~Cy_1 + R_1 sin(\theta_1+\omega_1 t),~Z_0^{(1)}\big)$ for $UAV_1$, and 

$\bigg(X_f^{(2)},Y_f^{(2)},Z_f^{(2)}\bigg) = \big(Cx_2 + R_2 cos(\theta_2+\omega_2 t),~Cy_2 + R_2 sin(\theta_2+\omega_2 t),~Z_0^{(2)}\big)$ for $UAV_2$.

The Euclidean distance between a UAV pair is called \textit{Link Distance} which is computed at time \textit{t} as,
\begin{equation}\label{Eq4}
    Link~Distance = \bigg(\big(X_f^{(1)}-X_f^{(2)}\big)^2 + \big(Y_f^{(1)}-Y_f^{(2)}\big)^2 + \big(Z_f^{(1)}-Z_f^{(2)}\big)^2 \bigg)^{\frac{1}{2}}
\end{equation}

Without the loss of generality, we assume the difference in the altitudes of the UAV pair is negligible as compared to their distance in the X and Y axis. Therefore, the term $\big(Z_f^{(1)}-Z_f^{(2)}\big)^2$ is dropped from \eqref{Eq4} for simplicity. Upon plugging the values of future coordinates of both UAVs, RHS of \eqref{Eq4} changes to, \\
$\bigg(\big(Cx_1 + R_1 cos(\theta_1 + \omega_1 t) - Cx_2 - R_2 cos(\theta_2 + \omega_2 t)\big)^2 + \big(Cy_1 + R_1 sin(\theta_1 + \omega_1 t) - Cy_2 - R_2 sin(\theta_2 + \omega_2 t)\big)^2\bigg)^{\frac{1}{2}}$.

It can be further simplified to,
\begin{equation}\label{Eq5}
\begin{split}
    \bigg(\big(Cx_1 - Cx_2\big)^2 &+ \big(Cy1 - Cy2\big)^2 + R_1^2 + R_2^2 - 2 R_1 R_2\big(cos(\theta_1 + \omega_1 t - \theta_2 - \omega_2 t)\big) \\
    &+ 2 R_1 \big((Cx_1 - Cx_2) cos(\theta_1 + \omega_1 t) + (Cy_1 - Cy_2) sin(\theta_1 + \omega_1 t)\big) \\&- 2 R_2 \big((Cx_1 - Cx_2) cos(\theta_2 + \omega_2 t) + (Cy_1 - Cy_2) sin(\theta_2 + \omega_2 t)\big)\bigg)^{\frac{1}{2}}
\end{split}
\end{equation}

The link between a UAV pair breaks when their $Link~Distance \geq Transmission~Range$. If $a = |Cx_1 - Cx_2|$, $sign_1 = \frac{(Cx_1 - Cx_2)}{a}$, and $b = |Cy_1 - Cy_2|$, $sign_2 = \frac{(Cy_1 - Cy_2)}{b}$, we can use $sin(\alpha) = \frac{b}{\sqrt{a^2+b^2}}$, $cos(\alpha) = \frac{a}{\sqrt{a^2+b^2}}$, and $\alpha = cos^{-1}\big(\frac{a}{\sqrt{a^2+b^2}}\big)$ to simplify \eqref{Eq4} to,
\begin{equation}\label{Eq6}
\begin{split}
    Link~Distance &= \big[a^2 + b^2 + R_1^2 + R_2^2 - 2R_1 R_2\big(cos(\theta_1 + \omega_1 t - \theta_2 - \omega_2 t)\big)\\
    &+ 2(sign_1) R_1 \sqrt{a^2 + b^2}~cos(\theta_1 + \omega_1 t - (sign_1*sign_2)\alpha)\\
    &- 2(sign_1) R_2 \sqrt{a^2 + b^2}~cos(\theta_2 + \omega_2 t - (sign_1*sign_2)\alpha)\big]^{\frac{1}{2}}
\end{split}
\end{equation}

Here, \eqref{Eq6} is a polynomial equation with one unknown variable \textit{t}.\\

\begin{figure}[!t]
	\centering
	\includegraphics[width=0.99\linewidth]{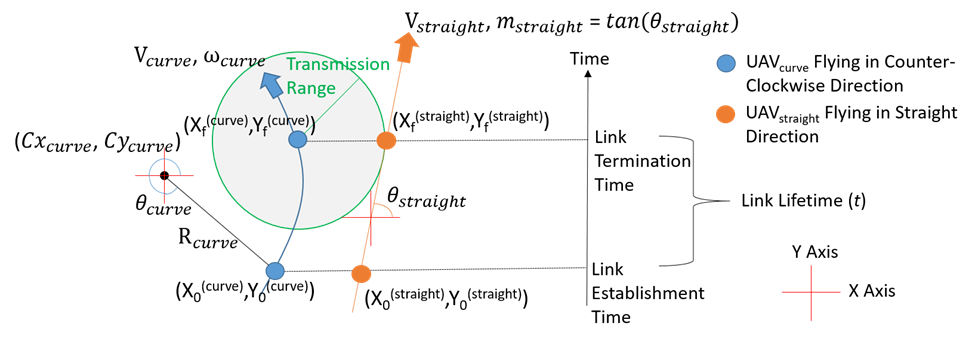}
	\caption {\textit{LLT} computation when one UAV flies in a curve while other UAV moves in a straight direction at an angle $\in$ [0, 2$\pi$] with respect to X-axis.}
	\label{Fig2}
\end{figure}

\noindent\textbf{Case B. One UAV flies in a curve and other UAV flies in straight direction at any angle $\in$ [0, 2$\pi$] with respect to X-axis}\par
Assume the current GPS locations for the pair $UAV_{staright}$-$UAV_{curve}$ in Fig. \ref{Fig2} are $(X_0^{(straight)}, Y_0^{(straight)})$ and $(X_0^{(curve)},Y_0{(curve)})$. For the UAV flying in a straight direction, let the speed be $V_{straight}$ and slope is $m_{straight}$. For the UAV flying in a curve, current trajectory center is $(Cx_{curve},Cy_{curve})$, radius is $R_{curve}$, velocity is $V_{curve}$, angular velocity is $\omega_{curve}$, initial displacement at the time of link establishment is $\theta_{curve}$, and movement direction is $Dir_{curve}$. Here, $Dir_{curve}$ is -1 for clockwise direction and +1 for counter-clockwise direction.

Using the approach discussed for case A, the polynomial equation for this case is,\\
\begin{equation}
\begin{split}
    Link~Distance &= \big[(V_{straight}*t)^2 + a^2 + b^2 + R_{curve}^2\\ 
    &+ 2(sign_1) R_{curve} \sqrt{a^2 + b^2}~\big(cos(\theta_{curve} + \omega_{curve}*t - (sign_1*sign_2)\alpha)\big)\\ 
    &- 2R_{curve}*V_{straight}*t~cos(\theta_{curve} + \omega_{curve}*t - tan^{-1}(m_{straight}))\\
    &- 2V_{straight}*t~cos(tan^{-1}(m_{straight}) - (sign_1*sign_2) \alpha)\big]^{\frac{1}{2}}
\end{split}
\end{equation}

Here, $a = \bigg|\big(Cx_{curve} - X_0^{(straight)}\big)\bigg|$, $sign_1 = \frac{\big(Cx_{curve} - X_0^{(straight)}\big)}{a}$, $b = \bigg|\big(Cy_{curve} - Y_0^{(straight)}\big)\bigg|$, $sign_2 = \frac{\big(Cy_{curve} - Y_0^{(straight)}\big)}{b}$, and $\alpha = cos^{-1}\big(\frac{a}{\sqrt{a^2+b^2}}\big)$.\\

\noindent\textbf{Case C. When both UAVs fly in a straight direction at random angles $\in$ [0,~2$\pi$] with respect to X-axis}\par
For the UAV pair, assume the current GPS locations are $(X_0^{(1)},Y_0^{(1)})$ and $(X_0^{(2)},Y_0^{(2)})$, slopes are $m_1$ and $m_2$, and velocities are $V_1$ and $V_2$, respectively. Using the approach discussed in case A, the polynomial equation obtained is,
\begin{equation}
\begin{split}
        Link~Distance &= \bigg[t^2 \bigg(V_1^2 + V_2^2 - 2V_1 V_2~cos\big(tan^{-1}(m_1) - tan^{-1}(m_2)\big)\bigg)\\
        &+ t\bigg(2V_1\big[(X_0^{(1)} - X_0^{(2)})~cos(tan^{-1}(m_1)) + (Y_0^{(1)} - Y_0^{(2)})~sin(tan^{-1}(m_1))\big]\\
        &-2V_2 \big[(X_0^{(1)} - X_0^{(2)})~cos(tan^{-1}(m_2)) + (Y_0^{(1)} - Y_0^{(2)})~sin(tan^{-1}(m_2))\big]\bigg)\\
        &+ \big(X_0^{(1)} - X_0^{(2)}\big)^2 + \big(Y_0^{(1)} - Y_0^{(2)}\big)^2 \bigg]^{\frac{1}{2}}
\end{split}
\end{equation}

\begin{figure}[!t]
	\centering
	\includegraphics[width=0.5\linewidth]{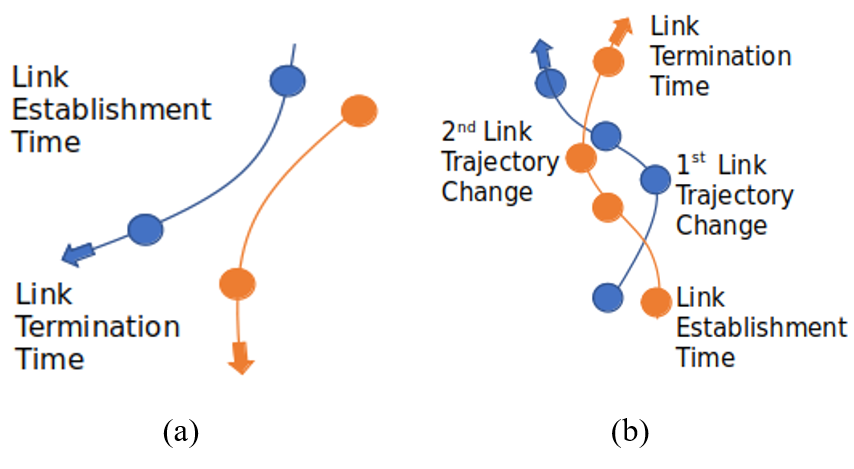}
	\caption {The link between a UAV pair can break (a) without any trajectory change (see left image) or (b) after multiple Link Trajectory Changes (see right image).}
	\label{Fig3}
\end{figure}

Note that the trigonometric functions in cases A and B are expanded up to 12 steps using Taylor’s expansion to get their close approximation. Then, the roots of the equation are computed using numerical method from which an appropriate root is selected that approximates the polynomial equation the best. We observed a very low approximation error in \textit{LLT} computation if the UAV trajectories do not change before the \textit{Link Termination Time} (see Fig. \ref{Fig3}(a) for example). Note that this approach can also be extended for elliptical UAV trajectory by using the minor and major axis of the elliptical trajectory.

\subsection{\textbf{\textit{LLT} Recomputation when UAV Trajectory Changes}}
\label{TrajectoryChange}
During the lifetime of a link, any UAV of a given UAV pair can change its trajectory. For example, both UAVs in Fig. \ref{Fig3}(b) change their respective trajectory once before the link termination (see $1^{st}$ and $2^{nd}$ Link Trajectory Change in Fig. \ref{Fig3}(b)). Since the trajectory change of a UAV may not be known a priori in an autonomous UAV network, the \textit{LLT} value should be reevaluated when either of the UAVs changes its trajectory before the link breaks. Therefore, a UAV pair calculates its \textit{LLT} when the link is first established and then updates the value whenever the UAV trajectory changes. This allows an accurate computation of \textit{LLT} value regardless of the past changes in the node trajectory.

\section{CONCLUSION}
\label{Conclusion}
Autonomous AN provides a low-cost solution to support a variety of applications in civil and military sectors because it can improve the network reliability and fault tolerance, reduce mission completion time through collaboration, and adapt to dynamic mission requirements. However, facilitating a seamless communication among nodes in such ANs is a challenging task due to their fast mobility, which results in the frequent link disruptions. Many existing AN-specific mobility-aware schemes restrictively assume that the airborne nodes fly in straight lines only, to reduce the uncertainty in the mobility pattern and simplify the calculation of \textit{LLT}. However, these schemes are not suitable for practical autonomous ANs, where an airborne node flies in a smooth trajectory, which includes mobility along the straight line as well as curves.

In this report, we described a mathematical framework to accurately compute the \textit{LLT} value of an airborne node pair, where each node flies independently in a randomly selected smooth trajectory. %We also discussed a mechanism by which a scheme can adapt to the random trajectory changes.

\ifCLASSOPTIONcaptionsoff
    \newpage
\fi

\end{document}